\documentclass[showpacs,preprintnumbers,amsmath,amssymb,12pt]{revtex4}
\usepackage{graphicx}
\usepackage{dcolumn}
\makeatletter
\newcommand{\doublespacing}{\let\CS=\@currsize\renewcommand{\baselinestretch}
{2.0}\tiny\CS}
\input{epsf}

\begin{document}
\title{ Spin-Orbit gauge and quantum spin Hall effect }

\author{B. Basu}
\email{banasri@isical.ac.in, Fax:91+(033)2577-3026}
\author{P. Bandyopadhyay}
\email{b_pratul@yahoo.co.in}
\affiliation{Physics and Applied Mathematics Unit\\
 Indian Statistical Institute\\
 Kolkata-700108 }
\vspace*{4cm}

\begin{abstract}
\begin{center}
{\bf Abstract}
\end{center}
We have shown that the non-Abelian spin-orbit gauge field strength of the Rashba and Dresselhaus interactions, when split into two Abelian field strengths, the Hamiltonian of the system can be re-expressed as a Landau level problem with a particular relation between the two coupling parameters. The quantum levels are created with up and down spins with opposite chirality and leads to the quantum spin Hall effect. 
\end{abstract}

\maketitle

\newpage
The surprising  prediction of spin Hall effect (SHE)i.e generation of a dissipationless quantum spin current perpendicular to the charge current at room tempearture in conventional hole-doped semiconductors \cite{3} has evoked a lot of interest in the investigation of $spin$ $physics$ in different contexts. In recent times, instead of charge of an electron, the electron spin is used for information procesing and storage and hence
the study of $spin$ related physics in semiconductors  have potential application in information industry and has become an emerging field of research in the area of condensed matter physics \cite{1,2}.  The SHE was found to be intrinsic in electron systems with substantial spin orbit (SO) coupling and the spin Hall conductance has a universal value \cite{4}. In a two dimensional electron gas with Rashba (RSO) and Dresselhaus (DSO)types of spin orbit interaction, the spin Hall conducatnce has a close relation to the Berry phase of conduction electrons \cite{5}. It is known that
the Rashba and Dresselhaus spin orbit interactions in a two dimensional semiconductor heterostructure can be treated in terms of the SU(2) non-Abelian gauge field
\cite{6}. The non-Abelian gauge field associated with these SO interactions gives rise to an effective magnetic field which is responsible for the rotation of the spins. The spin current arising out of these RSO and DSO have been studied in terms of the non-Abelian gauge field by several authors \cite{6,7}. Besides, quantum spin Hall effect(QSHE) is predicted in the conventional semiconductors in the absence of any external magnetic field
\cite{bernevig}. In semiconductors, the presence of strain gradient creates  the degenerate Landau levels by the spin -orbit coupling. 
In the present note, we predict fractional quantum spin Hall effect (FQSHE) in conventional semiconductors for a particular relation between the coupling strengths of RSO and DSO.
The Hamiltonian of a 2DEG with both types of spin orbit interaction  can be mapped to the Hamiltonian of a Landau-analogue problem with a particular constraint. The system segregates all up and down spins in two different layers. It is found  that the chiral states are with up spin and anti-chiral states are with spin down.  The spin Hall conductance is finite and qunatized in unit of $e/4\pi$ in each state. The  spin Hall conductivity can be computed in terms of the spin Berry phase acquired by the rotation of the spin eigen states in a closed path.

In two dimensional (2D) semiconductors, the spin-orbit interaction can be described in terms of two contributions to the model Hamiltonian. The standard Hamiltonian for the two dimensional sample with two types of spin orbit coupling is given by (in $c=\hbar=1$ unit)
\begin{equation}
H=H_0+H_R+H_D
 =\frac{k^2}{2m}+\alpha(k_x \sigma_y~-~k_y \sigma_x)~+~\beta(k_x \sigma_x~-~k_y \sigma_y)
 \end{equation}
 where as usual  ${k^2}={k_x}^2+{k_y}^2$ denotes the kinetic
energy term. The second term, namely the Rashba term \cite{rashba} $ {H_R}=\alpha({\sigma_x}{k_y}-
{\sigma_y}{k_x})$ comes due to the inversion asymmetry of the confining potential
 and the third term ${H_D}=\beta({\sigma_x}{k_x}-
{\sigma_y}{k_y})$ is the linear Dresselhaus coupling resulting due to the bulk
inversion asymmetry \cite{10}. The coefficient $\alpha$ is tunable by adjusting the external gate voltage and the coefficient $\beta$ is determined by the semiconductor material and the geometry of the sample. $\sigma_i$ are the usual Pauli
matrices.
It is known that the Hamiltonian $H$ 
 can be diagonalized exactly. The eigenstates are of the form
\begin{equation}\label{m1}
|\lambda~{\bf k}~>=\frac{1}{\sqrt 2}\left( \begin{array}{cc} e^{i\theta} \\
\lambda i \end{array}\right)
\end{equation}
where $\lambda=\pm 1$ is the band index and
$$\displaystyle{\tan\theta=\frac{\alpha k_y -\beta k_x}{\alpha k_x -\beta k_y}}$$
 The corresponding eigenenergy is given by
\begin{equation}\label{e1}
E_{\lambda \bf k}=\frac{k^2}{2m}-\lambda k \gamma (\phi)
\end{equation}
where 
\begin{equation}
\gamma (\phi)=\sqrt{\alpha^2+\beta^2-2\alpha\beta \sin~2\phi}
\end{equation}
with $\tan\phi=\frac{k_y}{k_x}$ and $k=|{\bf k}|$.
The band structure consists of two energy bands which are degenerate at ${\bf k}=0$.
The momentun states $|+~{\bf k}~>$ and $|-~{\bf k}~>$ with $\lambda=\pm 1$ are the spin eigenstates with 
up and down spin when $\alpha=\beta$ \cite{5} and the Berry phase for the transport of these states in k-space is zero.\\
 It is also observed \cite{6}that the Hamiltonian  (1) can be expressed in terms of an $SU(2)$ gauge potential as (in unit $c=\hbar=1$)
 \begin{equation}\label{h1}
 H^{R+D}=\frac{1}{2m}[{\bf k} -{\bf A}]^2 +\textrm{constant}
 \end{equation}
 This Hamiltonian is analogous to that of a charged particle in a uniform magnetic field.
 In this case, the
 effective vector potential ${\bf A}$ is spin dependent and the components are given by
 \begin{equation}
 A_x= m(\alpha \sigma_y + \beta \sigma_x),~~~~~~
 A_y=-m(\alpha \sigma_x + \beta \sigma_y)
 \end{equation}
 The effective $charge$ corresponding to the couplings are respectively given by
 $\theta=2m\alpha$ and $\bar{\theta}=2m\beta$ which have been absorbed in the SU(2) gauge potential ${\bf{A}}$. 
 In terms of this spin dependent gauge field in two dimensions, we can define a two component spin dependent field strength as 
 \begin{eqnarray}\label{f1}
 F_{xy}&=&\partial_x~A_y-\partial_y A_x+i[A_x,A_y] \nonumber \\
 &=& 2m^2(\alpha^2-\beta^2)\sigma_z  \nonumber \\
 &=& \left[ \begin{array}{cc}
 B_{z+} & 0 \\
0 & B_{z-}\end{array} \right]
 \end{eqnarray}
 Here, $B_{z+}=-B_{z-}=2m^2(\alpha^2-\beta^2)$ represents the $effective$ 
  magnetic field but correponds to an Abelian field strength. We may mention here that for $\alpha^2>\beta^2$, $B_{z+}$ is positive and $B_{z-}$ is negative. The situation is opposite for $\alpha^2<\beta^2$. It may also be noted  that the  magnetic field $B_{z\pm}$ in the $\pm z$-direction is equivalent to an electric field in the in-plane (xy) direction.

 The splitting of the non-Abelian gauge field strength  in (\ref{f1}) results to the Abelian gauge potential $\bf{{\cal {A}}}$ whose components satisfy the relation
 \begin{equation}\label{pot1}
 B_{z}=\partial_x {\cal{A}}_{y}-\partial_y {\cal{A}}_{x}
 \end{equation}
 The choice
of symmetric gauge enables us to write this Abelian vector potential as
 \begin{equation}\label{g2}
 {\bf{\cal{A}}}=\frac12(B_{z}(-y), B_{z}(x), 0)= m^2(\alpha^2-\beta^2)(-y,x,0)
 \end{equation}
 Let us  now construct a Hamiltonian  with the Abelian gauge potential as,
\begin{equation}\label{h3}
H=\frac{1}{2m}({\bf k}-{\bf{\cal{A}}}\sigma_z)^2
\end{equation}
where
\begin{equation}
{\bf{\cal{A}}}\sigma_z =\left[ \begin{array}{cc}  {\cal{A}}_+ & 0 \\
0& {\cal{A}}_- \end{array}\right]
\end{equation}
with ${\cal{A}}_+ =-{\cal{A}}_-$ and is given by eqn.(\ref{pot1}) and (\ref{g2}).
Our Hamiltonian (\ref{h3}) is exactly equivalent to the usual Hamiltonian of a charged particle in an uniform magnetic field, where the two different spin directions experience the opposite directions of the $effective$ $magnetic$ $field$.  Since,
  $[H,\sigma_z]=0$,  the  states are characterized by the spin in the z-direction. This system of 2DEG with an Abelian potential maps to harmonic oscillators when discrete energy levels are the Landau levels. The 
 Hamiltonian  (\ref{h3}) is expressed as 
 \begin{equation}\label{h4}
H=\left[ \begin{array}{cc}H_{\uparrow} & 0\\
0 & H_{\downarrow} \end{array} \right],\end{equation} where $H_{\uparrow}(H_{\downarrow})$  represents the Hamiltonian  of spin up(down) electrons and the corresponding gauge potentials are 
 ${\cal{A}}_{+}(x,y)$ and ${\cal{A}}_{-}(x,y)$ which are given by (\ref{g2})\\
 The explicit forms of the Hamiltonian $H_{\uparrow,\downarrow}$ is given by (omitting the suffix z)
  \begin{equation}\label{hh}
 H_{\uparrow,\downarrow}=\frac{1}{2m}[(k_x^2 +k_y^2)+ \frac{B^2}{4}(x^2+y^2)\mp B(k_xy-k_yx)]
 \end{equation}
 when ${\bf{\cal A}}=\frac{B}{2}(-y,x,0)$  and $B=2m^2(\alpha^2-\beta^2)$. We can check that for $\alpha^2>\beta^2$, $H_{\uparrow}$($H_{\downarrow}$) corresponds to the state with up spins (down spins). The scenario is opposite for the situation $\beta^2>\alpha^2$. 

 Now in an amazing way we can show that these states actually correspond to the Bloch states $|+ {\bf k}>$ or $|- {\bf k}>$ of our system  which is really interesting.\\
The Hamiltonian (1) or (\ref{h1})involving the non-Abelian gauge potential ${\bf A}$ can be recast into a Hamiltonian (\ref{h3}) 
with the Abelian gauge potential ${\cal{A}}$ for a specific spin direction provided the eigen energy value of both the Hamiltoians are identical. 

 The energy eigen value of the Hamiltonian (\ref{h3}) for any specific direction of spin is given by
 \begin{equation}\label{c1}
 E=(n+\frac{1}{2})\frac{B}{m}
 \end{equation}
 where $B=2m^2(\alpha^2-\beta^2)$.
 We have already mentioned that the energy corresponding to the Hamiltonian (1) is given by
 \begin{equation}\label{c2}
E_{\lambda \bf k}=\frac{k^2}{2m}-\lambda k \gamma (\phi)
\end{equation}
with $\lambda=\pm 1$. \\
For the lowest Landau level(n=0) and band index $\lambda=\pm 1$, both the eigen energies will equate if 
\begin{equation}\label{rel}
E_{\pm}=\frac{|B|}{2m}= \frac{k^2}{2m}\mp k \gamma (\phi)
\end{equation}
 where $B=2m^2(\alpha^2-\beta^2)$. We designate  $E_+$ for  $\alpha^2 > \beta^2$ and $E_-$ for   $\beta^2>\alpha^2$. Neglecting the terms of order O(4) of $\alpha$ and $\beta$, we find that the relation (\ref{rel}) is satisfied for both $E_+$ and $E_-$, with the constraint
 \begin{equation}\label{al}
 \alpha =\frac{4 \beta~k_x~k_y \pm \sqrt{16 \beta^2k_x^2k_y^2+2(k^2+m^2\beta^2)\frac{k^4}{m^2}}}{4(k^2+m^2\beta^2)}
 \end{equation}
 Eqn.(\ref{al}) infers that for a specific  heterostructure (i.e. for a fixed $\beta$), and for a wave vector ${\bf{k}}$ dependent $\alpha$,
the Hamiltonian (1) or (5) with the non-Abelian gauge potential $\bf{A}$ and the Hamiltonian (\ref{h3}) with the  Abelian gauge potential ${\cal{A}}$  can be made to have equivalent energy eigevalues. 
 It is presumed that the wave vector dependendent electric field ${\bf E}$ may generate the wave vector dependent Rashba coefficient.\\

 In this specific set up, with  $\alpha$, $\beta$ and ${\bf k}$ 
 satisfying  relation (\ref{rel},\ref{al}), it is found that the Hamiltonian (5)  involving the non-Abelian gauge field can be recast in the form of eqn. (10) which can be explicitly written in the form of eqn. (13).
The Hamiltonians $H_{\uparrow}$ and $H_{\downarrow}$ correspond to Hamiltonians of two opposite spin orientations. It is interesting to note that, in a conventinal semiconductor with both types of spin orbit coupling term a situation may arise which is analogous to that of Landau quantization. The Landau-like situation  is predicted in the absence of any external magnetic field or without the presence of any strain gradient \cite{bernevig}. 
  Our analysis suggets that a 2DEG with both RSO and DSO coupling strength will resemble a bilayer system where the two layers are placed together for a specific $\alpha-k$ relation with a fixed $\beta$ . From eqn. (13) we note that for $E_+$ (i.e.$\alpha^2>\beta^2$ )in one of the layers, there are spin up electrons in the presence of a up-magnetic field (chiral states)with positive charge Hall conductance quantized in unit of $(+e^2/\hbar)$. In the other layer we have electrons with down spin, and negative charge Hall conductance quantized in unit of $(-e^2/\hbar)$ experiencing a down-magnetic field (anti-chiral states). The two different spin states experience the opposite magnetic field. The LLL wave functions of the electrons in both the layers are given by the standard harmonic oscillator wavefunction with holomorphic and antiholomorphic coordinates respectively. Another interesting point to be noted here is that the  $|\uparrow>$ and $|\downarrow>$ states represents  the Bloch states $|+ {\bf k}>$ and $|- {\bf k}>$ respectively i.e. we have arrived at the positive and negative helicity states. For $E_-$(i.e. $\alpha^2<\beta^2$),  the whole scenario will be just reversed.  
  As the two layers are placed together, we will have finite spin conductance, though the total charge Hall conductuvity will be vanishing. The total spin Hall conductance will be quantized in unit of $\frac{e}{2\pi}$. 
   
 In analogy with the charge Hall conductivity, some characteristic features of  spin Hall conductivity may also be understood through the Chern-Simons Landau Ginzburg theory\cite{cs}. It is noted that the Hamiltonian describing the RSO and DSO coupling is T- invariant and P-breaking. However, the transformed Hamiltonian depicting the Landau level structure is both P- and T-invariant. Moreover, it is noted that within the bilayer the many body wave function \cite{bernevig}
 \begin{equation}
 \psi(u_i,w_i)=\prod_{i<j}(u_i-u_j)^m~\prod_{k<l}(w^*_k-w^*_l)^m~
 \prod_{r,s}(u_r-w^*_s)e^{-\frac 12(\sum_i u_iu^*_i+\sum_k w_kw^*_k)}
 \end{equation}
 is symmetric under the interchange of $u_i$, the up spin coordinate and $w^*_i$, the down-spin coordinate. This reflects the spin-up-chiral and spin-down-antichiral symmetry leading to the double Chern-Simons theory \cite{freedman}.  
 If the fields associated with the chiral and anti-chiral states (or the left and right movers of the theory) are given by 
 $f_{\mu}$ and $g_{\mu}$ then the low energy double Abelian Chern-Simons action of the spin Hall liquid is given by
 \begin{equation}\label{ac}
 S=\frac{\nu_+}{4\pi}~\int\epsilon^{\mu\nu\rho}f_\mu\partial_\nu f_\rho~
    - \frac{\nu_+}{4\pi}\int\epsilon^{\mu\nu\rho}g_\mu\partial_\nu g_\rho
 \end{equation}
  The Chern number $\nu_\pm$ is the filling factor of the corresponding states. As one can put a spin-up or spin-down electron in the system with same probablity we may consider $\nu_+= \nu_-$. This is the $U(1)_\nu \times {\overline{U(1)_\nu}}$ Chern Simos theory \cite{freedman}. The action (\ref{ac}) may be utilised to derive the fractional charge and statistics of the quaiparticles. The Berry phase  $\Gamma$ obtained by the rotation of the  eigenstates (chiral and anti-chiral) around a closed loop can be used to derive the filling factor $\nu$ through the relation \cite{hal}
  \begin{equation}
 \nu=\frac{\Gamma}{2\pi}
 \end{equation}
  which can be further employed to find the spin Hall conductivity of the system. But the  Berry phase $\Gamma$ obtained by rotation of the spin eigenstates (with vertical polarization)  around  a closed loop is found to be $\pm \pi$ \cite{bb,rel}, depending on  the direction of the effective magnetic field i.e. sign($\alpha^2-\beta^2$), which gives  $\nu=1/2$. Thus for $|\nu_+|=|\nu_-|$ and $\nu=|\nu_+| +|\nu_-|$ , we get the filling factor $|\nu_{\pm}| = \frac{1}{4}$. In the bilayer system, for the two layers placed together, the spin Hall conductivity is quantized in unit of $2\frac{e}{4\pi}$ \cite{bernevig}. Explicitly, we may write the spin Hall conductivity as,
  \begin{eqnarray}
  \sigma_{XY+}&=&2\frac{e}{4\pi}\times\frac{1}{4}=\frac{e}{8\pi}~~~~~~~~~~~~~~\rm{for}~~ \alpha^2>\beta^2 \nonumber \\
  \sigma_{XY-}&=&-2\frac{e}{4\pi}\times\frac{1}{4}=-\frac{e}{8\pi}~~~~~~~~~\rm{for}~~ \alpha^2<\beta^2
  \nonumber \\
  \sigma_{XY}&=& 0~~~~~~~~~~~~~~~~~~~~~~~~~~~~~~~\rm{for}~~~~\alpha^2=\beta^2
  \end{eqnarray}

 Finally, we intend to analyse the present framework in an external magnetic field. 
 Application of an external B-field to a system of 2DEG with both RSO and DSO couplings reveals a vertical spin eigenstate and the spin separation is acheived by both the SU(2) gauge asociated with RSO and DSO intearctions as well as the U(1) gauge related to the chirality of the external $B^{ext}$-field \cite{tan}.
 In our case, due to the splitting of the SU(2) gauge into $U(1) \times U(1)$ gauge, we see that for a particular condition eqn.(\ref{rel}     
 ), we have two spin eigenstates of vertically polarised spin current with opposite chirality.  It can be easily seen that the application of an external field will lead to a spin separation in such a way that we will have more (or less) up and down spin accumulation in the two layers in respect to the direction of the $B^{ext}$.
 The system with RSO and DSO couplings in an external $B^{ext}$-field such that 
 $B^{ext}=\nabla \times A^B$ may be represented by the Hamiltonian ($c=\hbar=1$) 
 \begin{equation}
 H^{SO+B}=\frac{1}{2m}(\Pi_x^2~+~\Pi_y^2 )+~\alpha(\Pi_x \sigma_y~-~\Pi_y \sigma_x)~+~\beta(\Pi_x \sigma_x~-~\Pi_y \sigma_y)
 \end{equation}
 where ${\bf{\Pi}}={\bf{p}}-e~{\bf A}^B$.
 This Hamiltonian can be reduced to the form
\begin{equation} H^{SO+B}=\frac{1}{2m}\left[(p_x-eA^B_x+\frac\theta2~\sigma_y+\frac{\bar{\theta}}{2}~\sigma_x)^2)+(p_y-eA^B_y+\frac\theta2~\sigma_x-\frac{\bar{\theta}}{2}~\sigma_y)^2\right]~+~\rm{constant}
 \end{equation}
 where $\theta=2m\alpha$ and $\bar{\theta}=2m\beta$ represent the effective charge associated with the RSO and DSO interactions.
 Comparing this with the Yang-Mills Hamiltonian
 \begin{equation} H^{YM}=\frac{1}{2m}\left[(p_x-{\bar{e}}{\bar{A}}_x)^2~+~+(p_y-{\bar{e}}{\bar{A}}_y)^2\right]~+~\rm{constant}
 \end{equation}
 where ${\bar{e}}$ is the coupling constant which we normalize as 1, we can identify
 \begin{eqnarray}
 \bar{A}_x =& eA_x^B-\frac\theta2~\sigma_y -\frac{\bar{\theta}}{2}~\sigma_x\\ \nonumber
 \bar{A}_y =& eA_y^B+\frac\theta2~\sigma_x -\frac{\bar{\theta}}{2}~\sigma_y
 \end{eqnarray}
 The effective field strength is given by
 \begin{eqnarray}
 {\bar{F}}_{xy}=&(\frac{\partial}{\partial x}\bar{A}_y~-
 ~\frac{\partial}{\partial y}\bar{A}_x)~+~[\bar{A}_x,\bar{A}_y]\\ \nonumber
 =&eB_z^{ext}+2m^2(\alpha^2-\beta^2)\sigma_z
 =&eB_z^{ext}+B_z^{R+D}
 \end{eqnarray}
 The explicit matrix form is
 \begin{eqnarray}
 \left[ \begin{array}{cc}
 {\bar{F}}_{xy}(+) & 0\\
 0 & {\bar{F}}_{xy}(-)\end{array} \right]& = &\left[ \begin{array}{cc}
 eB^{ext}_z+2m^2\alpha^2-2m^2\beta^2 & 0 \\
 0 & eB^{ext}_z-2m^2\alpha^2+2m^2\beta^2\end{array} \right] \\ \nonumber
 &=&\left[ \begin{array}{cc}
 eB^{ext}_z+B_z^{R+D}(+) & 0\\
 0 & eB^{ext}_z+B_z^{R+D}(-)\end{array} \right]
\end{eqnarray}
The effect of the external field in the two layers with up and down spins can be analysed from this expression.
We note that the external B-field with positive chirality will enhance the accumulation of up spins as this will reduce the effect of $B_z^{R+D}(-)$ having negative chirality for $\alpha^2>\beta^2$.
The situation is reversed when $B_z$ is introduced with opposite chirality as well as 
$\alpha^2<\beta^2$.

We  know  that a 2DEG with both Rashba and Dresselahus type of spin orbit interaction experience a non-Abelian spin dependent $effective $ magnetic force. On the other hand,
in presence of uniform magnetic field, the charged particles occupy orbits with discrete energy values, called  Landau levels which arise physically from velocity dependent Lorentz force. In our analysis, we have shown that when we split the non-Abelian SU(2) field strength into $U(1)_L \times U(1)_R $ i.e. into two Abelian field strengths then for a particular relation between the two spin orbit coupling strengths, the said Hamiltonian of  the 2DEG can be recast into a pair of Hamiltonians which are analogous to that of Landau level problem. It is intersting to note that without any external magnetic field, we can create Landau level  like discrete energy levels in two layers with opposite directional spin accumulation. The quantum Landau levels are created due to the $effective$ magnetic field in two opposite directions, present due to both types of SOC. As a result, we can observe quantum spin Hall effect with vanishing charge conductance but finite spin Hall conductivity. The application of an external magnetic field will enhance the accumulation of spin up and down electrons in both the layers.


\begin{thebibliography}{*99}
\bibitem{3} S. Murakami, N. Nagaosa and S.Zhang, Science {\bf 301}, 1348 (2003)
\bibitem{1}G. A. Prinz, Science {\bf 282}, 1660 (1998)
\bibitem{2} S. A. Wolf, Science {\bf 294}, 1488 (2001)
\bibitem{4} J. Sinova et al Phys. Rev. Lett. {\bf 92}, 126603 (2004)
\bibitem{5} S. Q. Shen, Phys. Rev. Lett. {\bf 92}, 126603 (2005)
\bibitem{6} N. Hatano et al., Phys.Rev.A {\bf 75}  032107  (2007)
\bibitem{7} S. H. Chen and C. R. Chang, Phys. Rev.B {\bf 77}, 045324 (2008) 
\bibitem{bernevig} B. A. Bernevig and S.C. Zhang, Phys. Rev. Lett. {\bf 96}, 106802 (2006)
\bibitem{rashba} E. I. Rashba, Sov. Phys. Solid State {\bf 2}, 1109 (1960); \\
~~~~~~~~~~Y.A. Bychkov and E.I. Rashba, J. Phys. C {\bf 17}, 6039 (1984)
\bibitem{10} G. Dresselhaus, Phys. Rev. {\bf 100}, 580 (1955)
\bibitem{cs} S.C. Zhang et al., Phys. Rev. Lett. {\bf 62}, 82 (1989)  
\bibitem{freedman} M. Freedman et.al., cond-mat/0307511 
\bibitem{hal} F.D.M. Haldane, Phys. Rev. Lett. {\bf 93}, 206602 (2004)
\bibitem{rel}  R. A. Bertlmann, K. Durstberger, Y. Hasegawa,B. C. Hersmayer,
 Phys. Rev. A {\bf 69}, 032112 (2004)
\bibitem{bb}  B. Basu, Europhys. Lett.{\bf 73}, 833 (2006);\\
~~~~~~~~~~~~ B. Basu and P. Bandyopadhyay, J. Phys. A Math and Gen. {\bf 4} 055301 (2008);\\
~~~~~~~~~~~~~~~B. Basu and P. Bandyopadhyay, Int. J. Geo. Methods of Mod. Phys. {\bf 41} 707 (2007)
\bibitem{tan} S. G.Tan et al., cond-mat/0701111    

\end{thebibliography}
\end{document}